\documentstyle[12pt] {article}
\title{POPULATION DYNAMICS OF CHILDREN AND ADOLESCENTS WITHOUT PROBLEMATIC BEHAVIOR}
\author{Vladan Pankovi\'c$^{1),2)}$, Nikola Vunduk$^{3)}$,
Milan Predojevi\'c $^{1),2)}$ \\
$^{1)}$Department of Physics, Faculty of Sciences, 21000 Novi
Sad \\ Trg Dositeja Obradovi\'ca 4., Serbia, vdpan@neobee.net\\
$^{2)}$Gimnazija, 22320 Indjija, Trg Slobode 2a., Serbia \\
$^{3)}$ Osnovna \v{s}kola "Petar Ko\v{c}i\'c", 22320 Indjija \\
Cara Du\v{s}ana 4., Serbia, nikola.v@neobee.net }

\begin {document}
\date{}
\maketitle

\vspace {0.5cm}

\begin {abstract}
In this work we suggest a simple mathematical model for the dynamics
of the population of children and adolescents without problematic
behavior (criminal activities etc.). This model represents a typical
population growth equation but with time dependent (linearly
decreasing) population growth coefficient. Given equation admits
definition of the half-life time of the non-problematic children
behavior as well as a criterion for estimation of the social
regulation of the children behavior.
\end {abstract}
\vspace {1cm}

It is well-known [1-3] that population of the children and
adolescents without problematic behavior (criminal activity, use of
the alcohol and narcotic drugs, etc.) has an interesting dynamics.
This population firstly (till some critical time moment) increases
and later (after given critical time moment) quickly decreases
during time. (Given population dynamics correlates to dynamics of
the positive influence of the parents and society at children and
adolescents.) In this work we shall suggest a simple mathematical
model for given population dynamics. This model represents a typical
population growth equation but with time dependent (linearly
decreasing) population growth coefficient. Given equation admits
definition of the half-life time of the non-problematic children
behavior as well as a criterion for estimation of the social
regulation of the children behavior.

So, we suggest the following first order differential equation
\begin {equation}
  \frac {d(x-c)}{dt} = (a-bt)(x-c)            .
\end {equation}
Here $x$ represents the population of the children (including
adolescents) without problematic behavior, $t$ - time moment from
childhood presented formally by an interval $[0,T]$  (here 0
formally denotes the beginning of the childhood while $T$ formally
denotes end of the childhood). Further, $a$ represents a positive
constant corresponding to influence of the positive factors (eg.
positive parents or, generally, social influence) at children
behavior, $b$ - positive constant corresponding to influence of the
negative factors at children behavior, and, $c$ - positive constant
corresponding to asymptotical limit of the children population
without problematic behavior (formally $x$ tends to $c$ when $t$
tends to infinity, but, of curse, $t$ is really smaller or equal to
$T$).

Obviously, equation (1) is similar to a typical equation of the
population growth, but in (1) instead of a constant population
growth coefficient there is a time dependent, i.e. linearly
decreasing term $(a-bt)$.

Equation (1) holds simple solution
\begin {equation}
   x = (x_{0}-c) \exp[\frac {at-bt^{2}}{2}] + c
\end {equation}
where $x_{0}> c$ represents the initial population of the children
without problematic behavior.

As it is not hard to see $x$ (2) has maximum
\begin {equation}
  x_{max} = (x_{0}-c) \exp[\frac{a^{2}}{2b}] + c
\end {equation}
for
\begin {equation}
  t = \frac {a}{b}
\end {equation}
that can be considered as a critical time moment. It means that $x$
increases for $t$ smaller than $\frac {a}{b}$ while $x$ decreases
for $t$ greater than $\frac {a}{b}$.

Finally, we can introduce half-life time of the non-problematic
children behavior, i.e. time moment
\begin {equation}
  t{\frac {1}{2}} = \frac {a}{b} [1 + (1 + \frac {b}{a}\ln 4)^{\frac {1}{2}}]
\end {equation}
that satisfies (2) and condition
\begin {equation}
  (x(t{\frac {1}{2}})-c) = \frac {x_{0}-c}{2}             .
\end {equation}
Then it can be stated that for $t{\frac {1}{2}}\geq T$ there is a
satisfactory social regulation of the children behavior, while in
opposite case, i.e. for $t{\frac {1}{2}}< T$, there is an
unsatisfactory social regulation of the children behavior.
Obviously, given statement can be considered as a criterion for
estimation of the social regulation of the children behavior.

In conclusion we can shortly repeat and point out the following. In
this work we suggested a simple mathematical model for the dynamics
of the population of children and adolescents without problematic
behavior (criminal activities etc.). This model represents a typical
population growth equation but with time dependent (linearly
decreasing) population growth coefficient. Given equation admits
definition of the half-life time of the non-problematic children
behavior as well as a criterion for estimation of the social
regulation of the children behavior.

\section {References}

\begin {itemize}

\item [[1]]  {\it  Handbook of prevention and treatment with children and adolescents: Intervention in the real world context}, eds. R. T. Ammerman, R. T. Hershen, (John Wiley and Sons, New York, 1997) and references therein
\item [[2]]  {\it  Serious and violent juvenile offenders: Risk factors and successful interventions}, eds. R. Loeber, D. Farrington (Sage Publications, Thousend Oaks, CA, 1998) and reference therein
\item [[3]]  {\it Handbook of antisocial behaviors}, eds. W. Stoff, J. Breiling, J. D. Masers (John Wiley and Sons, New York,1998) and references therein

\end {itemize}

\end {document}